\documentclass[notitlepage,onecolumn,prl,tightenlines,floatfix]{revtex4-2} 

\usepackage{geometry} 
\usepackage{graphicx}

\begin{document}

\title{Turn towards the crowd}
\author{Olivier Dauchot}%
\affiliation{Gulliver Lab, UMR CNRS 7083, PSL Research University, ESPCI Paris 10 rue Vauquelin, 75005 Paris, France}%
\date{21/05/2021} 

\begin{abstract}
{\it A type of polar self-propelled particle generates a torque that makes it naturally drawn to higher-density areas. The collective behaviour this induces in assemblies of particles constitutes a new form of phase separation in active fluids.}
\end{abstract}

\maketitle
Polar active fluids consist of self-propelled particles with a broken head–tail symmetry that imbues them with a polarity. Being inherently out of equilibrium, they borrow energy from their environment to convert it to translational motion in the direction of this polarity. As in regular fluids, the particles may interact through some attractive or repulsive pairwise potential, but their polarity can also induce interactions in the form of alignment rules. Depending on the density and the details of this interaction, energy taken up on the microscopic scale can be converted into essentially two types of macroscopic collective behaviour, namely, collective motion (see Figure 1a) and motility-induced phase separation (see Figure 1b). And now, writing in Nature Physics, Jie Zhang and colleagues have revealed a new mechanism for phase separation in assemblies of Janus particles, which has its roots in a specific type of orientational interaction [1]. \\

This story of polar active fluids is a little more complicated than a simple conversion of microscopic energy. First, the transition to collective motion is discontinuous and actually takes place via a microphase separation into large propagating bands [2]. Second, the agents are likely to slow down with the local density due to a crowding effect. As a result, a population of non-aligning, purely repulsive agents may exhibit a so-called motility-induced phase separation into a dense aggregate, surrounded by a low-density gas [3]. In the presence of alignment, this condensation, and the associated slowing down of the particles, can actually hinder collective motion [4]. Conversely, alignment can either suppress or promote [5, 6] standard motility-induced phase separation in a population of repulsive agents. 
The phase-separation mechanism that Zhang et al discovered does not require the slowing down of the particles with the local density. Instead, the electrophoretic Janus particles, the team studied experimentally, present a stronger repulsion on the rear than on the front and thereby produce non-reciprocal torques that reorient the particle motion toward high-density regions. Particles thus self-propel up their own density gradient — an example of ‘autotaxis’ — and this leads to phase separation (see Figure 1c). \\

\begin{figure}
\includegraphics[width = 0.5\textwidth]{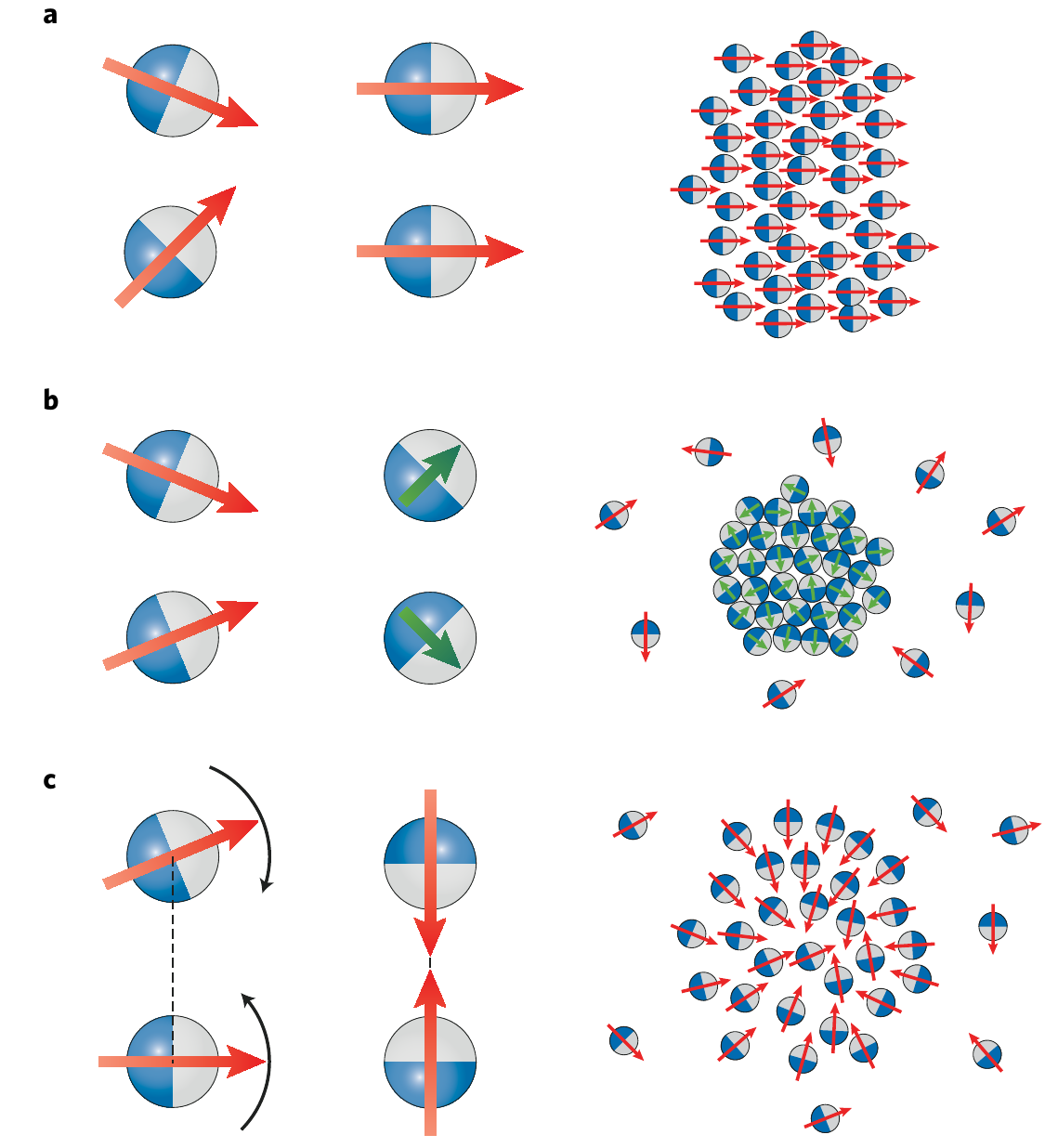}
\caption{{\bf Types of collective behaviour in polar active fluids.} Janus particles with broken head–tail symmetry (illustrated here in blue and white) have a nominal velocity (red arrows) that can either reorient without slowing down, or slow down (green arrows) without aligning. a, Alignment without slowing down can induce collective motion. b, Slowing down without alignment can result in motility-induce phase separation. c, A density-dependent torque without slowing down can give rise to autotactic phase separation, as shown by Zhang et al [1].}
\vspace{-0.5cm}
\end{figure}

The resulting aggregate remains fluid, as opposed to the dense clusters typically obtained in repulsion-based motility-induced phase separation. Also, they display substantial ordering of the polarity field, yet no ordering of the velocities. Finally, they exhibit a fast population turnover, with particles leaving and entering the aggregates at a high rate. From a biological perspective, such properties could favour group functions, such as efficient exchange of information between the inside and the outside of the aggregate, as well as among aggregates.\\

The types of phase and collective phenomenon one might expect from a set of microscopic rules can be determined using standard tools of out-of-equilibrium statistical physics. By averaging out the irrelevant degrees of freedom, one obtains dynamical equations for the large-scale fields of interest, here the polarity and the density fields. The coupling between these fields effectively decides the types of phase and instability the system will exhibit. \\

In the ordered flocking phase, the inherent fluctuations of the polarity field, associated with its rotational symmetry, carry and amplify the density fluctuations. These lead to the so-called giant fluctuations, a clear signature of the out-of-equilibrium nature of the flocking phase [7]. In the case of the motility-induced phase separation, the scalar coupling between the motility (the amplitude of the velocity field) and the density field leads to a positive self-trapping feedback, responsible for the phase separation. In their case, Zhang et al were able to demonstrate theoretically that a vectorial coupling between the velocity field and the density gradient, the auto-taxis, is responsible for the phase separation.\\

Exploring model experimental systems, such as the one introduced here, has proven to be a good strategy for probing mechanisms for the onset of collective behaviours in active systems [5, 8, 9]. Once identified, they call for further theoretical investigations, as several open issues remain. For example, one may wonder whether it is truly a bulk phase separation taking place or rather a microphase separation, as suggested by the presence of persistent boundaries between merging aggregates. Another question of interest is the level of universality of these scenarios. Very little is known about the critical properties — both static and dynamic — that should develop at the tip of the coexistence regime.  Stochastic hydrodynamics and renormalization group techniques [2, 7] offer a promising, though challenging, route to address such fundamental questions.\\

References:
{\small 
\begin{itemize}
\item [1] Zhang, J., Alert, R., Yan, J., Wingreen, N. S. and Granick, S. Nat. Phys. (2021).
\item [2] Solon, A. P., Chaté, H. and Tailleur, J. From Phase to Microphase Separation in Flocking Models: The Essential Role of Nonequilibrium Fluctuations. Phys. Rev. Lett. 114, 068101 (2015).
\item [3] Cates, M. E. and Tailleur, J. Motility-Induced Phase Separation. Annual Review of Condensed Matter Physics 6, 219–244 (2015).
\item [4] Geyer, D., Martin, D., Tailleur, J. and Bartolo, D. Freezing a Flock: Motility-Induced Phase Separation in Polar Active Liquids. Phys. Rev. X 9, 031043 (2019).
\item [5] Van der Linden, M. N., Alexander, L. C., Aarts, D. G. A. L. and Dauchot, O. Interrupted Motility Induced Phase Separation in Aligning Active Colloids. Phys. Rev. Lett. 123, 098001 (2019)
\item [6] Großmann, R., Aranson, I. S. and Peruani, F. A particle-field approach bridges phase separation and collective motion in active matter. Nat. Commun. 11, 5365 (2020).
\item [7] Toner, J. and Tu, Y. Long-range order in a two-dimensional dynamical XY model: How birds fly together. Phys. Rev. Lett. 75, 4326–4329 (1995).
\item [8] Bricard, A., Caussin, J.-B., Desreumaux, N., Dauchot, O. and Bartolo, D. Emergence of macroscopic directed motion in populations of motile colloids. Nature 503, 95–98 (2013).
\item [9] Geyer, D., Morin, A. and Bartolo, D. Sounds and hydrodynamics of polar active fluids. Nature Materials 17, 789–793 (2018).
\end{itemize}}

\end{document}